%% file: main.tex
\documentclass[acmsmall]{acmart}

\usepackage{algorithmic}
\usepackage{graphicx}
\usepackage{textcomp}
\usepackage{xcolor}
\usepackage{blindtext}
\usepackage{multirow}
\usepackage{subfigure}
\usepackage{tcolorbox}
\usepackage{enumitem}
\usepackage{array}
\usepackage{rotating}
\usepackage{fancyhdr}
\usepackage{csquotes}
\usepackage{dcolumn}
\usepackage{booktabs}
\usepackage{caption}
\usepackage{colortbl}
\usepackage{svg}
\usepackage{amsmath}
\usepackage{bigstrut}
\newcolumntype{d}[1]{D{.}{.}{#1}}

\AtBeginDocument{%
  \providecommand\BibTeX{{%
    \normalfont B\kern-0.5em{\scshape i\kern-0.25em b}\kern-0.8em\TeX}}}

\setcopyright{acmcopyright}
\copyrightyear{2023}
\acmYear{2023}
\acmDOI{XXXXXXX.XXXXXXX}

\acmConference[CSCW]{XXXX}{XXXX}{XXXX}
\acmPrice{15.00}
\acmISBN{978-1-4503-XXXX-X/18/06}

\begin{document}

\title[Characterizing Usability Issue Discussions in OSS Projects]{Characterizing Usability Issue Discussions in Open Source Software Projects}

\author{Arghavan Sanei}
\email{Arghavan.sanei@polymtl.ca}
\author{Jinghui Cheng}
\email{Jinghui.cheng@polymtl.ca}
\affiliation{%
  \institution{Polytechnique Montreal University}
  \city{Montreal}
  \state{Quebec}
  \country{Canada}
}

\renewcommand{\shortauthors}{Sanei and Cheng}

\begin{abstract}
Usability is a crucial factor but one of the most neglected concerns in open source software (OSS). While far from an ideal approach, a common practice that OSS communities adopt to collaboratively address usability is through discussions on issue tracking systems (ITSs). However, there is little knowledge about the extent to which OSS community members engage in usability issue discussions, the aspects of usability they frequently target, and the characteristics of their collaboration around usability issue discussions. This knowledge is important for providing practical recommendations and research directions to better support OSS communities in addressing this important topic and improve OSS usability in general. To help achieve this goal, we performed an extensive empirical study on issues discussed in five popular OSS applications: three data science notebook projects (Jupyter Lab, Google Colab, and CoCalc) and two code editor projects (VSCode and Atom). Our results indicated that while usability issues are extensively discussed in the OSS projects, their scope tended to be limited to efficiency and aesthetics. Additionally, these issues are more frequently posted by experienced community members and display distinguishable characteristics, such as involving more visual communication and more participants. Our results provide important implications that can inform the OSS practitioners to better engage the community in usability issue discussion and shed light on future research efforts toward collaboration techniques and tools for discussing niche topics in diverse communities, such as the usability issues in the OSS context.
\end{abstract}



\maketitle

\section{Introduction}
\input{s_Introduction}

\section{Related work}
\input{s_RelatedWork}

\section{Methods}
\input{s_Methods}

\section{Results}
\input{s_Results_RQ1}
\input{s_Results_RQ2}
\input{s_Results_RQ3}
\input{s_Results_RQ4}

\section{Discussion}
\input{s_Discussion}
\input{s_Limitations}

\section{Conclusion}
\input{s_Conclusion}

\bibliographystyle{ACM-Reference-Format}
\bibliography{references}

\appendix

\end{document}

%% file: s_Introduction.tex
Open source software (OSS) development is a unique collaborative endeavor that involves various types of stakeholders who assume diverse, often overlapping roles (e.g., maintainers, developers, designers, and end users)~\cite{Cheng2019CHASE}. The nature of being ``open'' allows people with any background to use and contribute to OSS projects. But, at the same time, the diverse users' characteristics and needs are often difficult to consolidate and be adequately addressed by an OSS community. To make the situation worse, a typical OSS community usually values technical contributions and functional complexity of the system and does not have the relevant expertise to address concerns related to UI/UX design~\cite{wang2020argulens, Wang2022IEEESoftware}. As a result, while other software qualities (e.g., functionality, defects, security, and performance) are important, usability (as an attribute broadly capturing the ability of the software in allowing users to complete relevant tasks in an effective, efficient, safe, and satisfiable fashion~\cite{nielsen1994usability}) became a distinctive issue for OSS and is among the most neglected areas of OSS development and research~\cite{nichols2006usability,raza2010improvement, Hellman2021Facilitating}. Many OSS projects still suffer from inferior usability, making it difficult for the OSS model to reach its full potential. Because of these factors, OSS usability has recently attracted avid interest from both the research community~\cite{wang2020argulens, Hellman2021Facilitating, nichols2006usability,rajanen2015power} and the practitioners (e.g., the Open Source Design initiative\footnote{https://opensourcedesign.net}). 

Currently, a common practice adopted by many OSS communities when collaboratively addressing usability is through issue tracking systems (ITSs)~\cite{Wang2022IEEESoftware, cheng2018open}, such as GitHub Issues\footnote{https://docs.github.com/en/issues/tracking-your-work-with-issues}. ITSs are collaborative platforms that allow any member of an OSS community to report and discuss topics related to the product and the development process. They do not only help the software maintenance team keep track of the concerns raised by the community and the tasks to be completed, but they also support community building, allowing members to provide mutual support and contribute to the OSS projects. The discussion of usability issues in the ITSs usually involves various types of stakeholders and allows users and designers to be directly connected with the developers and maintainers.

However, using current ITSs to address OSS usability is far from an ideal approach. Usability issues are often buried in a large number of other topics in ITSs, making them difficult to be discerned. OSS developers and maintainers also tend to ignore usability issues or put them on a lower-priority list~\cite{Wang2022IEEESoftware}. Plus, ITSs are traditionally developer-centric tools; thus, end users and designers often find them intimidating to use~\cite{Hellman2021Facilitating, hellman2022characterizing}. Nonetheless, ITSs are still the main venue in which users' concerns can be integrated into the OSS development process. Because of these factors, it is important to understand the characteristics of usability issues that are currently discussed in ITSs in order to identify promising practical recommendations and research directions to better support OSS communities in addressing usability. Such knowledge is not established in the literature. Consequently, in this paper, we address this goal through an empirical study of the usability issue discussions on GitHub, one of the most widely used OSS project hosting platforms.

Particularly, we targeted five popular OSS projects hosted on GitHub: three data science notebook projects (Jupyter Lab, Google Colab, and CoCalc) and two code editor projects (VSCode and Atom). All the selected projects are under active development, have a large and diverse user base, and maintain a large number of issues discussing various topics. We collected the issue discussion data on these five projects to form our dataset. In accordance with our overall goal, we formulated four main research questions that guided our study, which we present below, along with their motivations, our approach to answering them, as well as a summary of the main findings.

\textbf{RQ1: What central concerns discussed in ITSs are related to usability?}

The ITSs are used to discuss a vast variety of issues related to OSS projects. Usability issues are often buried among a large number of discussions about different topics, making the community lose focus on this important aspect. Thus, identifying the usability-related project concerns will serve as a first step to understanding the current OSS communities' interests; this knowledge will, in turn, help distill opportunities that cater to these interests and address the corresponding blind spots. To this end, we performed topic modeling on the entire issue dataset to identify the prominent concerns targeted in the issues and the ones related to usability. We found a total of seven topics in issues from the data science notebook projects (two are related to usability) and 19 topics from the code editor projects (five are related to usability). The usability concerns, according to these topics, are mainly associated with visual design and user interaction mechanisms.

\textbf{RQ2: What is the frequency of usability issues discussed in OSS projects, and what aspects of usability are frequently targeted in those issues?}

While previous studies have established ITS as an important tool for OSS usability discussion, there is very little knowledge about how frequently this tool is used for this purpose. The topic modeling results do not provide an accurate estimate of usability versus non-usability issues because of the limitations of automated methods. It also only gives us a coarse view of the community concerns at the project level and does not touch on different aspects of usability. We thus pose this RQ to further understand the extent to which ITS is used for the discussion of usability in general and of different usability aspects in particular; we used Nielsen's heuristics~\cite{nielsen2005ten} as a categorization of usability aspects because of its widespread acceptance and established influence. To answer this RQ, we selected representative samples of each project from our dataset (a total of 1,734 issues) and performed a manual analysis to determine if each sampled issue was related to usability, and if yes, the usability aspect captured by Nielsen's heuristics it targeted. We found that the ratio of usability issues in ITSs varied across the projects, ranging from 9.9\% to 25.1\%, with an average of 17.4\%. The most frequently discussed usability aspects are \textit{flexibility and efficiency of use} and \textit{aesthetic and minimalist design}, constituting more than 60\% of all usability issues (36.4\% and 27.5\%, respectively).

\textbf{RQ3: What are the distinctive characteristics of the usability issues compared to non-usability issues?}

Usability issues touch on unique concerns of an OSS project, particularly since they are often reported from the users' perspective rather than the system's perspective. Thus we hypothesized that usability issues in ITSs display different characteristics than non-usability issues. Identifying these characteristics can further help distinguish usability issues from the large number of issues discussed in ITSs. To this end, we conducted statistical analyses using various metrics related to the issue reports (e.g., whether an issue is posted with visual support and the number of community reactions to the issue report) and the issue discussion threads (e.g., discussion length and the number of discussion participants). We found that, when compared to non-usability issues, usability issues were reported with more visual content and tended to receive a larger number of emoji reactions from the community. In certain projects (Jupyter Lab and VSCode), usability issues also involved more participants than non-usability issues, who took longer to discuss and reach a closure.

\textbf{RQ4: What are the experiences and backgrounds of people who posted usability issues?}

In addition to the characteristics of the issues themselves, we hypothesized that the community members who posted the usability issues might have different characteristics. To test this hypothesis, we conducted statistical analysis on the experiences and backgrounds of three groups of issue posters: (1) users who only posted usability issues, (2) users who only posted non-usability issues, and (3) users who posted both usability and non-usability issues. We found that people who have posted both usability and non-usability issues (groups 3) were generally more active and had more experiences on GitHub than users of the two other groups.

Overall, our study yielded important insights into the extent to which usability issues are discussed in OSS issue tracking systems, the major usability aspects covered in those issues, and the distinctive characteristics of usability issues and the corresponding reporters. Our results provided actionable implications for OSS practitioners to better address usability issues and shed light on future research efforts toward collaboration techniques and tools for better discussing and managing niche topics in diverse communities, such as the usability issues in the OSS context.

%% file: s_RelatedWork.tex
Our work is related to previous studies that focused on (1) open source software issue discussions, (2) the usability of open source software, and (3) collaboration in software development. We briefly review each body of literature in this section.

\subsection{OSS issue discussions}
One of the crucial tools for open source software (OSS) communities is issue tracking systems (ITSs). This tool assists OSS community members in completing various activities related to the development and maintenance of OSS. There is a wide variety of previous investigations focusing on approaches, techniques, and methods to leverage the presented information in these tools. These previous efforts focused on various practical applications using information from the ITSs, including analysis of design topics~\cite{viviani2018design}, retrieval and analysis of requirements~\cite{heck2017framework, morales2019speech}, triaging bugs~\cite{Xia2017}, analyzing impacts of changes \cite{huang2017packages}, predicting the type of software defect\cite{patil2020predicting}, to name a few.

Apart from leveraging information for software engineering tasks, a group of studies concentrated on issue discussion threads and explored the roles of ITSs in supporting diverse software stakeholders to work collaboratively. For example, Bertram et al.~\cite{bertram2010communication} investigated the use of ITSs in software development and pinpointed the role of issue discussion threads as the main conversation channel among software team members. Arya et al.~\cite{arya2019analysis} also discovered sixteen types of embedded information in issue discussion threads. Further, three patterns of conversation (i.e., collaboration, feedback, and monolog) used by stockholders while discussing various topics in ITSs were identified by Rath et al.~\cite{rath2020request}. Our investigation is built upon this body of literature to explore how usability issues are discussed in ITSs.

\subsection{OSS usability}
Addressing usability in OSS projects is a challenging task. Due to the developers' insufficient knowledge of users' requirements, inadequate experience in design, and an extensive focus on features and complexity of the systems, the usability of OSS projects has been frequently neglected~\cite{nichols2006usability, Wang2022IEEESoftware}. Schwartz et al.~\cite{schwartz2009integrating} identified that usability in OSS is often an afterthought issue, addressed late in the project's life-cycle; they argued that OSS needs the help of the HCI community to improve its user interaction design.

Currently, one frequently used tool for collecting the community members' requirements and feedback relevant to usability issues is ITSs. This type of platform has been used to collaboratively report, discuss, address, and argue various aspects of usability topics~\cite{nichols2006usability, bach2009floss, iivari2011participatory, Wang2022IEEESoftware}. Although ITSs are practical platforms to involve community members, they suffer from a group of limitations when used for discussing usability issues~\cite{Yusop2017}. 
For example, these issues are often greatly impacted by the personal opinions and the experiences of the participants~\cite{cheng2018open}. Furthermore, the community members who participate in ITSs usability discussions usually experience an information overload, which impacts how effective their participation is in these communications~\cite{Baysal2014}. Wang et al.~\cite{wang2020argulens} attempted to address this issue by suggesting an argumentation model for extracting and consolidating the different opinions voiced by community members in ITSs. Overall, getting involved in OSS usability discussions is still a challenging task, deferring OSS developers, end-users, and user experience experts to collaboratively make meaningful contributions to improve OSS usability~\cite{bach2009designers, iivari2013configuring, rajanen2015power}. Building on top of this body of literature, our study systematically characterizes how usability issues are currently discussed in open source issue tracking systems in the hope of identifying opportunities and providing implications toward a new type of collaborative platform for addressing OSS usability.

\subsection{Collaboration in software development}
Software development is innately a collaborative endeavor~\cite{bertram2010communication}. The usability issues usually involve the collaboration among three types of software stakeholders, namely developers, designers, and end users. In the following, we discuss the previous studies that investigated the collaboration among these three roles in software development.

The challenges of developer-designer collaboration have long been investigated in the literature~\cite{Brown2012}. These challenges usually originate from the misunderstanding and the different mindsets of the two parties~\cite{Li2017}. The tension is sometimes influenced by the conflict between the need for fast development iteration (e.g., in Agile) and integration of UI/UX~\cite{Ferreira2007AgileDI}. Several previous works have focused on methods and techniques to ease the integration of UI/UX tasks in the fast-paced iteration process ~\cite{chamberlain2006towards, Lundstrom2015, Leiva2019}. For example, Leiva et al.~\cite{Leiva2019} investigated tools for supporting developer-designer collaboration when transitioning the design into working prototypes. The recent focus on standardized design languages, such as Design Systems~\cite{Churchill2019, lamine2022understanding}, also serves as an effort to alleviate these challenges.

At the same time, the involvement of end users in the software development process is a crucial theme in many CSCW and HCI studies~\cite{Kensing1998PD, bach2010involving}. This involvement is somewhat facilitated by the ``openness'' of the OSS projects~\cite{terry2010perceptions, nichols2006usability, Hellman2021Facilitating}. On the one hand, end-users may provide important feedback directly to the OSS development teams~\cite{raza2010improvement}. For example, Ko et al.~\cite{ko2010powerusers} explored the impacts of power users in OSS projects. On the other hand, obstacles such as code-centric perspectives and inadequate supporting tools constantly hinder OSS members and contributors from taking users into the OSS development and maintenance loop~\cite{Hellman2021Facilitating}. Moreover, the OSS discussion environments are not always welcoming. Negative interactions such as toxicity and incivility can appear and deter people, particularly end-users, from participating in collaborative efforts~\cite{li2021code,ferreira2021shut}. Built upon this body of literature, in this paper, we investigate how OSS stakeholders collaborate in usability issue discussions.

%% file: s_Methods.tex
In this section, we describe our data collection, preprocessing, and sampling approach, as well as the analysis methods used to answer our research questions. The generated datasets and the analysis code are available as a replication package\footnote{https://github.com/HCDLab/UsabilityIssuesSupplementaryMaterial}.

\subsection{Projects Selection}
In our research, we focused on projects hosted on GitHub, which is considered one of the biggest OSS hosting platforms. We targeted well-known and active OSS applications that include a Graphical User Interface (GUI) so that they may attract diverse and active community members to discuss usability issues. Because both authors are active users of data science notebooks and code editors, we targeted these two application domains to ensure proper understanding and analysis of the project issues.

Using the aforementioned criteria, we selected three data science notebook projects that are popular among students and scientists to be applied for data analysis and modeling: (1) \textbf{Jupyter Lab}\footnote{https://github.com/jupyterlab/jupyterlab}, which has been developed and maintained not only as the next-generation user interface for Jupyter projects but also as an extensible environment for interactive data science based on the Jupyter Notebook. (2) \textbf{Google Colab}\footnote{https://github.com/googlecolab/colabtools}, a cloud-based data science notebook platform that can be used without requiring any local Jupyter notebook environment setups. (3) \textbf{CoCalc}\footnote{https://github.com/sagemathinc/cocalc} or Collaborative Calculation, which is another cloud-based data science platform that has a special focus on education; in addition to a typical online notebook, this project provides course management features such as assignments management and whiteboard discussion spaces.

We also chose two popular text and code editor projects for comparison: (1) \textbf{VSCode}\footnote{https://github.com/microsoft/vscode} and (2) \textbf{Atom}\footnote{https://github.com/atom/atom}. Both projects are lightweight, customizable, and extendable. With the support of community-built extensions, they support various features similar to IDEs, such as code highlighting, code completion, and debugging.

All the selected OSS projects have sophisticated user interaction designs and involve collaborators, contributors and users with diverse backgrounds (e.g., data scientists, machine learning experts, developers, students, and amateurs who focus on data analysis). These factors can benefit us by providing rich and varied characteristics of usability issues and discussions. 

\subsection{Data Collection}\label{DataCollection}
After selecting the projects, we gathered the data from their GitHub repositories. The data collection process was done in June and July 2021, using the GitHub REST API (https://docs.github.com/en/rest). We collected all the issues in these repositories with their comments and corresponding commits. After gathering data, we extracted issues in the \textit{closed status}; we focused on closed issues, which allowed us to understand the entire life-cycle of the issue discussion threads. A total of \textit{133,229} are closed issues from 139,948 collected. 

In our analysis, we only focused on issues that are written in English. To this end, we performed a preliminary analysis on a random sample of 383 issues (95\% confidence level and $\pm5\%$ confidence interval from all the closed issues) to examine the performance of several automated language detection techniques (including SpaCy FastLang\footnote{https://spacy.io/universe/project/spacy\_fastlang}, Compact Language Detector v2\footnote{https://github.com/CLD2Owners/cld2}, Compact Language Detector v3\footnote{https://github.com/google/cld3}, and Google Translate API\footnote{https://cloud.google.com/translate/docs/basic/detecting-language}). We found that the language detection feature in the Google Translate API achieved a superior agreement with our manual language labeling on the sampled issues ($\kappa=0.99$). We thus used this API on the entire dataset to remove non-English issues; particularly, we used the Googletrans Python packaging of the API\footnote{(https://pypi.org/project/googletrans/}. After this step, a total of \textit{127,282} English issues remained in our dataset (\textit{7,348} in the data science notebook projects and \textit{119,934} in the code editor projects).

It is worth noting that in our dataset, there were issues that addressed community or development process matters instead of the software itself; there were also ones that were even considered irrelevant to the project. However, those issues were essentially non-usability and still added to the challenges of identifying and addressing usability issues for the OSS communities. We thus intentionally kept them to understand the true proportion and characteristics of usability issues.

\subsection{Data Preprocessing and Sampling}\label{DataPreprocessingSampling}
For the topic modeling and quantitative analysis methods described below in Sections ~\ref{sec:method_TopicModeling} and~\ref{sec:QuantitativeAnalysis}, we performed natural language processing techniques on our issue discussion dataset. To ensure the accuracy and performance of these techniques, we prepossessed the natural language issue discussion data (formatted with Markdown syntax). First, we removed textual contents that are listed in Table~\ref{table:preprocess}. Then for topic modeling, we also followed Wang et al.'s ~\cite{wang2019does} recommendations to remove the stop words and performed lemmatization of nouns, verbs, adjectives, and adverbs. For other quantitative analyses, we kept stop-words and did not perform tokenization, lemmatization, or stemming because these tasks are done by the tools we used (see Section~\ref{sec:QuantitativeAnalysis}).

\begin{table}[t]
\small
\centering
\caption{Textual contents removed during pre-processing.}
\begin{tabular}{p{4cm}p{6cm}}
\hline
\multicolumn{1}{c}{\textbf{Content Removed}} & \multicolumn{1}{c}{\textbf{RegEx used to detect the content}} \\ \hline
Code snippets & \verb|```|, \verb|```python| \\
HTML tags & \verb|<!--.*?-->|, \verb|<[^>]*>| 
\\
URLs & \verb|https/S+|, \verb|http/S+| \\
Reference quotation & \verb|>/S+|, \verb|<.*?>|\\
Mentioning a user & \verb|(@[A-Za-z0-9]+)| \\
System paths & \verb|c:/S+|, \verb|e:/S+|, etc. \\
Frequently used entity names & Project names (e.g., \verb|VSCode|, \verb|CoCalc|, etc.), browsers' names (e.g., \verb|Chrome|, \verb|Firefox|, etc.)\\
Issue templates & We checked the recommended templates for issue posting in each project, then removed the template texts (e.g., \verb|Steps to Reproduce|, \verb|### Prerequisites|, \verb|**Describe the expected behavior**|, etc.) \\
\hline
\end{tabular}
\label{table:preprocess}
\end{table}

Moreover, our RQ2 is answered through a manual content analysis of the dataset, and RQ3 and RQ4 are based on this manual labeling. To manage the manual effort in this analysis, we constructed a representative sample of the closed English issues in each repository, considering a 95\% confidence level and a $\pm5\%$ confidence interval~\cite{sim1999StatisticalInference}. Table~\ref{table:Summary-datasets} summarizes the sample sizes. In total, the sampled dataset included 1,734 issues, of which 976 are from the data science notebook projects and 758 are from the other code editors; an average of 346.8 issues were sampled from each project. Subsequent quantitative analysis was then done on each project sample. Instead of creating a representative sample of the entire dataset, this sampling and analysis strategy is adopted because VSCode has a substantially larger amount of issues than the other projects, extremely overshadowing the results if sampling and analysis are done on the entire dataset.

\begin{table*}[t]
\small
\centering
\caption{Summary of our datasets and sampled data}
\begin{tabular}{c|lccc}
\hline
\centering
\textbf{Type} & \textbf{Project}  & \textbf{\#Closed issues}  
& \textbf{\#Closed \& English issues} & \textbf{\#Sampled}\\ 
\hline
\multirow{4}{*}{\begin{sideways}Notebooks\end{sideways}}
& Google Colab &      1,246           
& 1,181  & 294  \\
& CoCalc              &      2,390         
&      2,193   & 331 \\ 
& Jupyter Lab         & 4,080                 
& 3,974   & 351       \\
& \textbf{Total}  & \textbf{7,716}   
& \textbf{7,348} & \textbf{976} \\
\hline
\multirow{3}{*}{\begin{sideways}Editors\end{sideways}}
& Atom                &    15,446         
& 12,000      & 375       \\
& VsCode              & 110,067    
& 107,934 & 383 \\

&  \textbf{Total} & \textbf{125,513}  
& \textbf{119,934} & \textbf{758} \\
 \hline       
\end{tabular}
\label{table:Summary-datasets}
\end{table*}

\subsection{Topic Modeling} \label{sec:method_TopicModeling}
To answer \textbf{RQ1 - what central concerns discussed in ITSs are related to usability}, we applied a topic modeling approach to cluster and distinguish the list of topics in the entire dataset of the selected projects. Because of the differences in the nature of the projects and the size of the datasets, we aimed to build two topic models for data science notebook projects and code editor projects, respectively. Particularly, we used the Latent Dirichlet Allocation (LDA)~\cite{Blei2003} technique as the topic modeling algorithm; the LDA technique was applied separately to the preprocessed datasets of 7,348 closed English issues in data science notebook projects and 119,934 issues in code editor projects (see Section~\ref{DataCollection} and Section~\ref{DataPreprocessingSampling}). We built the topic models using the MALLET toolkit~\cite{McCallumMALLET}, which implements the Gibbs sampling algorithms~\cite{Geman1984} for LDA. We used the default hyperparameters ($\alpha\_sum=5.0$, $\beta=0.01$) and used both uni-gram and bi-grams in the LDA model since previous research found that this approach would improve the model quality~\cite{Tan2002, Openja2020}.

The LDA technique requires a predetermined number of topics $N$ as input. In order to find the optimal number of topics for each topic model, we varied $N$ ranging from 2 to 50 in unit increments and used the \textit{coherence score} as the measure of the model quality. The coherence score evaluates the semantic relation of the top-most terms in each topic~\cite{newman2010automatic}; it thus represents the interpretability and understandability of the topics~\cite{chang2009reading}.

Once built, the topic models offered the keywords for each identified topic as well as different probabilities of topics for each datapoint (i.e., issue post); we considered the highest probability as the dominant topic for each issue. In order to interpret and label the topics, we selected the top ten issues with the highest dominant probabilities for each topic. We then used the topic keywords and the ten selected issues to label each topic. Particularly, two authors first performed the labeling independently. Then they discussed in meetings to address disagreements and determine the final topic labels. Finally, the two authors discussed determining if each topic is related to usability or not by examining the topic label, as well as the keywords and dominant issues.

\subsection{Qualitative Analysis} \label{sec:Qualitative Analysis}
To answer \textbf{RQ2 - what is the frequency of usability issues discussed in OSS projects, and what aspects of usability are frequently targeted in those issues}, we conducted a qualitative content analysis ~\cite{krippendorff2018content} on the sampled dataset by adopting the following steps. Firstly, we labeled each sampled issue with either \textit{usability} or \textit{non-usability}. Usability is traditionally defined by various attributes related to the user interaction design of the projects, such as learnability, efficiency, memorability, errors, and satisfaction~\cite{nielsen1994usability}. We considered these attributes when labeling the usability issues. Secondly, for each usability issue, we identified the main usability aspect touched by the issue by classifying the issue with one of the \textit{ten Nielsen heuristics}~\cite{nielsen2005ten}; we considered an issue is related to a heuristic if the problem reported in the issue violates the heuristic or the solution proposed aligns with the heuristic. There were a few cases in which the issue could be classified with more than one heuristic; in those cases, we considered the most dominant and closely related heuristic to label the issue. For the above two steps, two authors first conducted the coding independently, then their agreement was assessed with Cohen's Kappa~\cite{Viera2005Kappa}. The agreement levels for the usability/non-usability coding and the usability aspect labelling were both considered `Near Perfect' ($\kappa=0.94$ and $\kappa=0.82$, respectively). The two coders then discussed their coding and reached a full agreement. These inter-rater agreements indicated that considering a single, most related heuristic for each issue is a proper choice.

Apart from distinguishing usability issues and labeling their usability aspects with Nielsen's heuristics, we also investigated how the usability aspects were manifested in the issue discussions. To this end, we conducted an open coding ~\cite{sandelowski1995qualitative} on the usability issues. The open coding focuses on identifying the factors in the usability aspects (i.e., the heuristics) that were discussed in the issues. For this step, two authors first conducted the open coding individually. Then, they held multiple meetings to discuss and merge their coding results.

\subsection{Quantitative Analysis} \label{sec:QuantitativeAnalysis}
To answer RQ3 and RQ4, we conducted a series of quantitative statistical analyses on the sampled dataset of each project. The purpose of these analyses is to investigate the correlational relationship between the independent and dependent variables that we focused on; in other words, we do not aim to claim casual relationships. Below, we describe the analysis methods for answering each research question.

\subsubsection{Characteristics of usability issues}
\label{sec:rq3-methods}
To answer \textbf{RQ3 - what are the distinctive characteristics of the usability issues compared to non-usability issues}, we considered two independent variables: (1) the issue type (i.e., usability vs. non-usability) and (2) the usability aspect (captured by Nielsen's heuristics) touched by the issue. We compared the groups of each independent variable on the following metrics as dependent variables. These metrics are derived from prior research~\cite{Abdalkareem2020, Sanei2021impacts, Agrawal_2022visual} and are grouped into two categories. First, we considered the following four metrics that characterize the issue reports:
\begin{itemize}
    \item \textbf{Reporting issues with visual support}: This is a binary metric indicating whether the issue report is purely textual or it includes visual content such as images and videos.
    \item \textbf{Number of reactions to the report}: GitHub has the feature of allowing community members to make emoji reactions (e.g., thumb up, thumb down, smiley face, heart, etc.) to the issues. This metric counts the number of reactions to the issue report to indicate how much attention the community pays to the issue.
    \item \textbf{Sentiment of issue post}: This metric indicates the emotional polarity of the issue post, categorizing the post into having a \textit{positive}, \textit{neutral}, or \textit{negative} sentiment. We used Senti4SD~\cite{calefo20018sentiment} to identify the sentiment of issue posts; the \textit{Senti4SD} technique is found to have a superior performance in detecting sentiment in GitHub issue discussions~\cite{Novielli2021}.
    \item \textbf{Tone of issue post}: This metric captures the emotional category of the issue post. We used \textit{IBM Watson Tone analyzer}~\cite{yin2017tone} as the tone identification tool. This tool classifies natural-language text into seven different tones: \textit{sad}, \textit{frustrated}, \textit{sympathetic}, \textit{impolite}, \textit{polite}, \textit{satisfied}, and \textit{excited}.
\end{itemize}

For the sentiment analysis and the tone analysis, we used the configurations of previous research~\cite{Sanei2021impacts} to achieve a satisfactory level of reliability and ensure the performance of the identification. Then, we examined the following metrics that characterize the issue discussion threads:
\begin{itemize}
    \item \textbf{Discussion length}: This metric indicates the complexity of the discussion by calculating the number of comments included in the issue discussion thread.
    \item \textbf{Number of participants}: This metric represents the community engagement in the issue discussion by calculating the number of unique users who participated in the issue thread.
    \item \textbf{Time to first comment}, \textbf{time to last comment}, and \textbf{time to close issue}: These metrics represent the discussion timeline by capturing, respectively, (1) how fast the community responds to the issue post, (2) how long it took for the community to discuss the issue temporally, and (3) how long it took for the project maintainer to determine issue closure. These metrics were determined by calculating the timestamp difference (in seconds) between the issue post and (1) the first comment, (2) the last comment, and (3) the event of issue closure, respectively.
\end{itemize}

Among the above metrics, three of them (i.e., reporting issues with visual support, as well as sentiment and tone of the issue post) are nominal data, while the rest are all interval data. For the nominal metrics, we used a chi-squared test to analyze the differences among the groups of each independent variable; then, for the usability aspects, if a significant result was found, we conducted posthoc pairwise comparisons with Holm-Bonferroni correction to identify the aspect pairs contributing to the difference. For the interval metrics, Shapiro-Wilk tests indicated that all are normally distributed. Thus we used the ANOVA test to explore the difference among the groups of each independent variable. If a significant difference was found in the usability aspects, we further conducted pairwise posthoc analysis using t-tests with Holm-Bonferroni correction to identify the independent group pairs contributing to the difference. The alpha level in all the tests was set at 0.05. 

\subsubsection{Experiences and backgrounds of usability issue posters} \label{sec:rq2-methods}
To answer \textbf{RQ4 - what are the experiences and backgrounds of people who posted usability issues}, we first categorized all issue posters in our dataset into three groups: (1) users who only posted usability issues, (2) users who only posted non-usability issues, and (3) users who posted both usability and non-usability issues. We considered this category of issue posters as the independent variable. We compared these three groups on five metrics as dependent variables representing the experiences and backgrounds of these issue posters; the selection of the metrics is inspired by previous research~\cite{Abdalkareem2020, Sanei2021impacts}. These five metrics are:
\begin{itemize}
    \item \textbf{Number of posted issues}: This metric is used to indicate how active the users contribute to the issue reports. It is calculated by counting all the issues posted by each unique user ID in our entire dataset.
    \item \textbf{Number of posted comments}: This metric is used to indicate how active the users participate in the issue discussion processes. It is calculated by counting all the comments posted by each unique user ID in our entire dataset.
    \item \textbf{Number of years on GitHub}: This metric indicates how experienced the users are in using and contributing to open source projects on the GitHub platform. It is a field included in each issue poster's profile.
    \item \textbf{Number of repositories owned}: This metric indicates how experienced the users are in managing open source projects. It is a field included in each issue poster's profile.
    \item \textbf{User's role in the project}: This metric indicates the degree of involvement and the type of contribution the users make to the projects. Users on GitHub projects usually have three types of roles: (1) \textit{Members} and \textit{Collaborators} are part of the organization and have certain privileges in the project, (2) \textit{Contributors} are users who made code contributions to a project but do not have special privileges, and (3) \textit{Others} who never made code contributions and only participated in issue discussions. It is a field included in each issue poster's profile concerning each project.
\end{itemize}

For the first four metrics (interval data type), Shapiro-Wilk tests indicated that all are normally distributed. Thus we used the ANOVA test to explore the difference among the three independent groups. If a significant difference was found, we further conducted pairwise posthoc analysis using t-tests with Holm-Bonferroni correction to identify the independent group pairs contributing to the difference. For the fifth metric (nominal data type), we used a chi-squared test to analyze the difference among the three groups; then, if a significant result was found, we conducted posthoc pairwise comparisons with Holm-Bonferroni correction. The alpha level in all the tests was, again, set at 0.05.

%% file: s_Results_RQ1.tex
\subsection{RQ1: Central concerns discussed in ITSs and their connection to usability}
\begin{table*}[t]
\small
\centering
\caption{Topics identified in the notebook and the code editor projects. (\textit{Bold font indicates usability-related topics.})}
\resizebox{\textwidth}{!}{
\begin{tabular}{l|p{4cm}p{6.5cm}p{6cm}}
\hline
Pro.    &  Topic                   
& Some topic words & Definition \\
\hline
\multirow{8}{*}{\begin{sideways}Notebook projects\end{sideways}}
& N0- Installation/build bugs  &  npm, packages, install, anaconda, lib, error, module, build, version, dependency & Reporting bugs during installation, packaging, and/or building processes. \\\bigstrut

&   N1- Developing and supporting extensions & extensions, create, packages, document, widget, master, pull, release, documentation, docs &  Focusing on developing the extensions or documenting the project to facilitate extension development. \\\bigstrut

& N2- Errors related to coding &  questions, drive, issue, web, browser, error, fail, usage, use, GPU, Tensorflow &  Discussing issues happen while using the application for coding. \\\bigstrut

&   \textbf{N3- Visual rendering issues} & images, content, theme, screen, dark, cell, block, plot, line, icons, color  &   The usability issues which are results of text rendering and component displaying problems. \\\bigstrut

& N4- Communication tools & project, create, sage, course, page, chat, user, time, student, worksheet, link  & Extensions, add-ons, or features for users to communicate through the software. \\\bigstrut

&   N5- Environment issues & file, kernel, browser, directory, terminal, server, path, console, environment, command & Describe issues that occurred on the python kernels that the application's host. \\\bigstrut

& N6- \textbf{User interaction issues} & cell, tab, editor, menu, line, markdown, click, button, view, panel, keyboard, cursor, scroll & Reporting issues linked to the user interface and user interaction design of the notebook application. \\
\hline

\multirow{8}{*}{\begin{sideways}Code editor projects\end{sideways}}

& C0- Code syntax highlighting and auto-completion  &  code, file, HTML, syntax, typescript, IntelliSense, language, highlighting, javascript, snippet &  Syntax highlighting and code completion problems in different programming languages. \\\bigstrut

& \textbf{C1- Visual design} & design, visible, icon, gif, image, shows, display, markdown, hover, menu, button, size & Usability issues related to the visual presentation and aesthetics of the user interface. \\\bigstrut

& C2- Installation bugs/crashes &  installation, setup, runtime, compiler, services, exception, trace, thrown, handle, threading   &  Bugs/crashes during the process of installing and/or setting up the application.\\\bigstrut

& \textbf{C3- Editor layout } &  editor, click, view, tab, menu, bar, focus, explorer, panel, context, markdown, tree, pane, icon  & Usability issues related to the layouts of different windows and panels in the code editor. \\\bigstrut

& \textbf{C4- Accessibility} &  color, accessibility, keyboard, clipboard, navigate, toolbar, accessible, insider, disabled, expected  &   Focusing on the practice of making the project usable for people with disabilities. \\\bigstrut

&  C5- System failure  &  debugging, crashing, hacking, mode, following, described, functionality, provides, crashed, failed   &   Freezing and crashing of the system without any helpful information or message. \\\bigstrut

& C6- Performance when startup &  start, resources, performance, startup, share, connection, sources, service, console, renderer   & Reporting issues related to slow startup of the application. \\\bigstrut

& C7- Release engineering issues  &  extensions, reproduce, disabled, installing, update, ubuntu, insiders, build, setting  & Issues related to application release, including packaging, versioning, and cross-platform support. \\\bigstrut

& \textbf{C8- Code editing interaction} &  line, search, cursor, press, select, editor, word, enter, copy, shortcut, replace   &   Enhancements to code editing experiences, such as code navigation, selection, and modification. \\\bigstrut

&  C9- Terminal/console issues &  terminal, command, launch, integrated, remote, build, debug, shell, process, bash, console, failed  & Problems linked to terminal, shell, or bash scripting, commanding, and/or launching. \\\bigstrut

& C10- Building error &  node, modules, error, npm, failure, build, code, typescript, server, load, dependency, import  &   Problems connected to the process of building the source code of the application. \\\bigstrut

& C12- Service crash &  code, service, status, CPU, server, truncated, memory, crash, Linux, unity, restart, times   &   Reporting crashes of external services used by the application.\\\bigstrut

&   C13- Update/upgrade problem &   update, core, thrown, uncaught, fail, upgrade, undefined, log, browser, updater, reinstall  & Issues related to failing in updating and upgrading the application. \\\bigstrut

&  \textbf{C14- Theme-related issues}  &  color, theme, scroll, font, dark, background, mouse, mode, width, light, hover, scrollbar   &  Issues related to the color, font, and other visual aspects of the user interface. \\\bigstrut

&   C15- Compilation error &  browser, common, uncaught, base, check, debug, typeerror, preferences, control, master, identifier  & Problems related to compiling the application project from its source code. \\\bigstrut

& C16- Debugging issues &  debug, time, console, debugging, breakpoint, process, helper, restart, log, slow, hit  &   Issues related to the debugging features offered by the applications. \\\bigstrut

&   C17- Crash on macOS X &  framework, crashing, electron, apple, kernel, foundation, frame, application, tools, mac  & Problems related to running the software on macOS. \\\bigstrut

&  C18- Extension development  &  extension, settings, feature, request, support, api, time, allow, case, json, provide   &   Issues linked to developing extensions to the application. \\\bigstrut

&   C19- File management &  file, folder, git, workspace, save, path, directory, root, control, commit, branch, repo, repository  &  Issues concentrated on versioning, managing, and accessing files or folders.   \\
\hline
\end{tabular}
}
\label{table:Summary-topics}
\end{table*}

To understand the OSS communities' central concerns exposed in the issues, we conducted topic modeling (LDA) analyses on the notebook and code editor issue datasets. The best topic model for the notebook projects included seven topics (\textit{coherence score = 0.51}), while that for the code editor projects included 20 topics (\textit{coherence score = 0.66}). When we labeled the topics, topic 11 of the code editor projects was removed because it is related to the management process of one particular project; this resulted in 19 topics for the code editor projects. Table~\ref{table:Summary-topics} summarizes these topics. After topic labeling, we identified two topics in the notebook projects and five topics in the code editor projects that are related to usability, as indicated in bold font in Table~\ref{table:Summary-topics}.

Overall, we found a broad range of topics related to usability and non-usability aspects in data science notebook and code editor projects. There are several similarities among the topics identified in those two types of projects. Particularly for usability-related topics, Topic N3 (visual rendering issues) is related to Topics C1 (visual design) and C14 (theme-related issues); and Topic N6 (user interaction issues) is related to Topics C3 (editor layout) and C8 (code editing interaction). For topics not related to usability, both types of projects included topics related to bugs/crashes during installation and build (N0, C2, C7, C10, and C15), support for developing extensions (N1 and C18), and coding and debugging issues (N2, C0, and C16).

There are several unique topics discussed only in one type of project. For usability-related topics, accessibility issues are only discussed in code editor projects (C4), particularly in VSCode. For topics not related to usability, environment issues (N5) are unique for notebook projects, which typically provide a python kernel; plus, issues related to communication tools (N4) only appeared in notebook projects, which offer such features. On the other hand, code editor projects have unique topics related to system failure and crashing (C5, C12, and C17), performance (C6), terminal/console (C9), updating (C13), and file and version management (C19).

\vspace{4pt}
\begin{tcolorbox}[colframe=black,colback=gray!10,boxrule=0.5pt,arc=.3em,boxsep=-1mm]
\textbf{RQ1 Main Findings}: 
Seven out of the 26 topics identified from our analysis were related to usability. Notebook and code editor projects have similar usability-related topics that touch on visual design and user interaction mechanisms. Specific usability topics, such as accessibility, are also found but only discussed in certain projects.
\end{tcolorbox}

%% file: s_Results_RQ2.tex
\subsection{RQ2: Frequency and aspects of usability issues}\label{sub: descriptive-characteristics}

\begin{figure}[t]
    \centering
    \includegraphics[width=.6\textwidth]{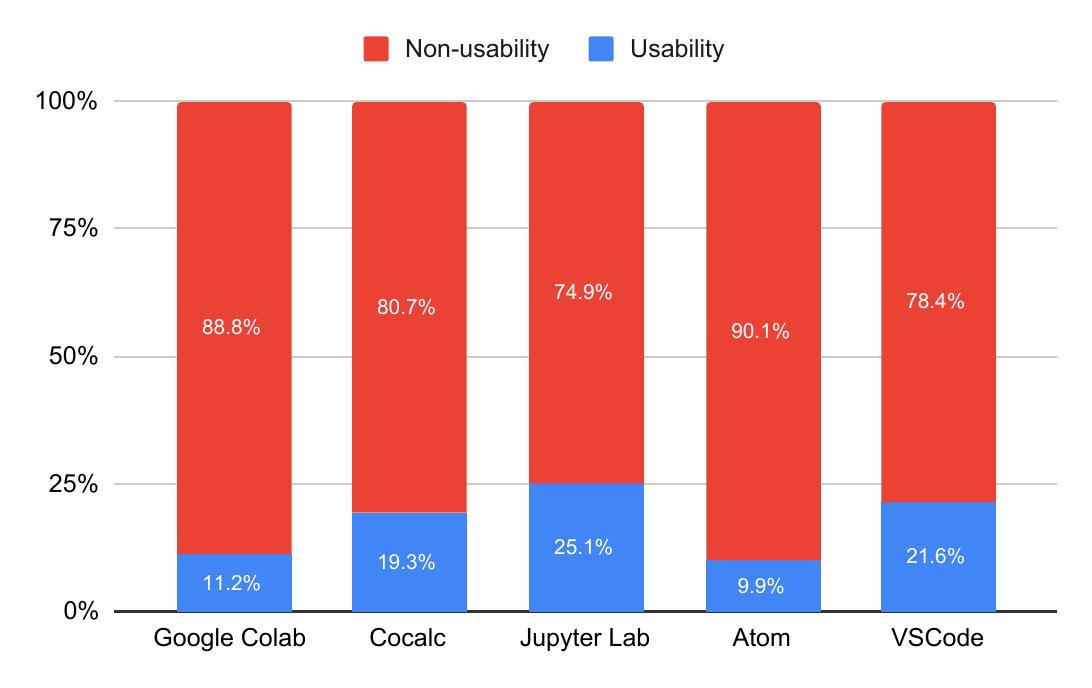}
    \caption{Frequency of usability and non-usability issues in each project}
    \label{fig:frequency-usability}
\end{figure}

While topic modeling gives us an overview of general topics discussed in the ITS, the results are very coarse and can be inaccurate. To identify the extent to which the ITS is used for usability issue discussion and to understand the usability aspects frequently focused on by the OSS communities, we conducted a manual content analysis on a representative sample of issues from each analyzed project. We found that across our analyzed projects, an average of 17.4\% ($SD=6.6\%$) issues in the samples are related to usability (a total of 305 issues in the entire sample of 1,734 issues). This percentage of usability issues varied depending on the projects (see Figure~\ref{fig:frequency-usability}). Particularly, the Jupyter Lab project participants paid more attention to usability issues (25.1\% in our sample). On the other hand, in Google Colab and Atom, the percentages of usability issues are only around 10\% (11.2\% and 9.9\%, respectively).

\begin{figure}[t]
    \centering
    \includegraphics[width=\textwidth]{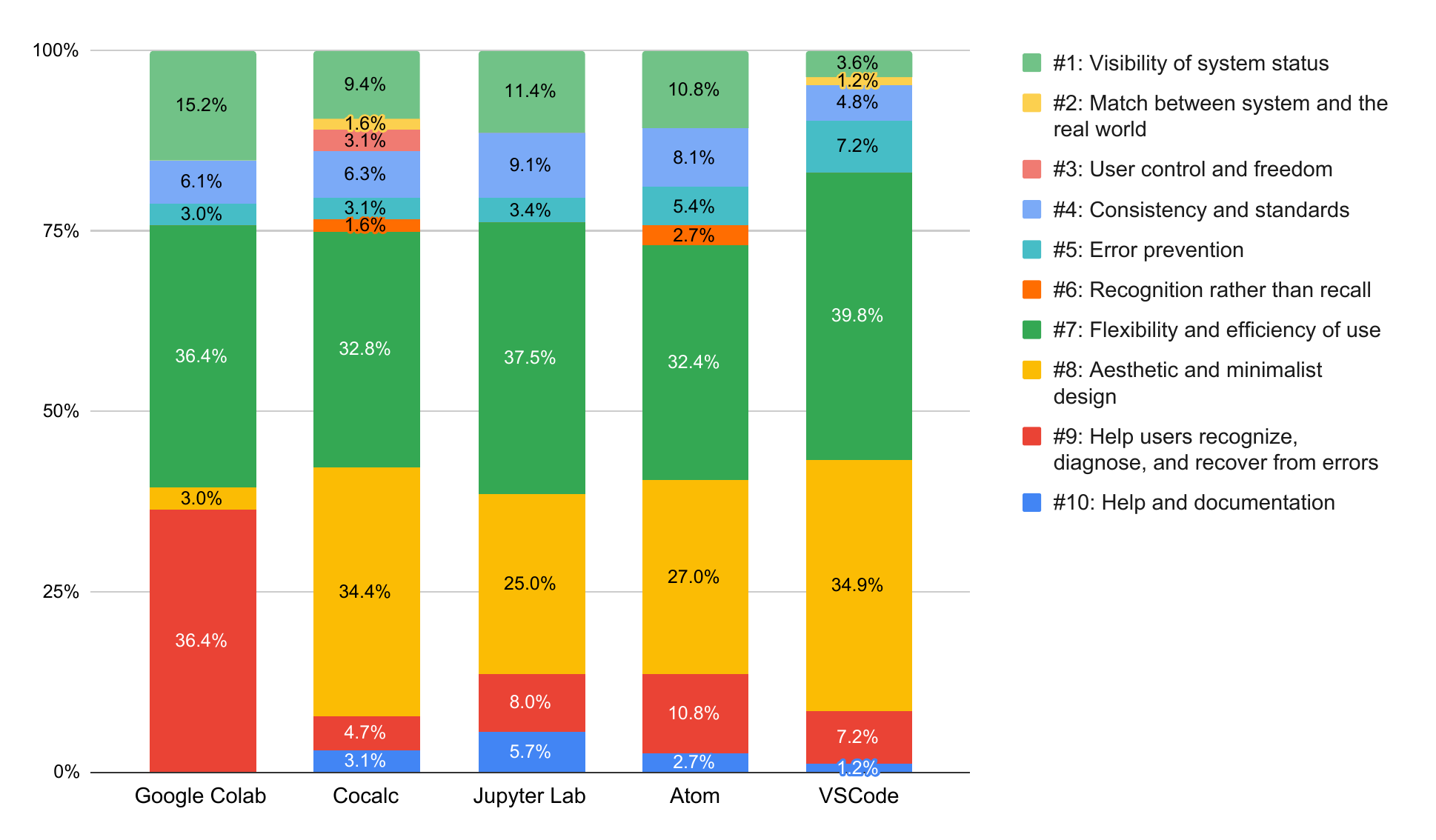}
    \caption{Frequency of Nielsen's usability heuristics in usability issues in each project}
    \label{fig:frequency-heuristics}
\end{figure}

After determining an issue as related to usability, we specified the usability aspect targeted in each issue leveraging Nielsen's heuristics~\cite{nielsen2005ten}. Figure~\ref{fig:frequency-heuristics} summarizes the frequency of each usability aspect (i.e., heuristic) that appeared in the usability issues of each project. Overall, there are six most frequently targeted aspects in the selected projects that we discuss below, along with the themes on the focus of discussion in each aspect that we identified from open coding.

\textbf{\#7: Flexibility and efficiency of use} (a total of 111 issues, or 36.4\% in the entire sample). 
    Issues coded with this usability aspect in our dataset focus on features or enhancements for speeding up the interaction for the convenience of both inexperienced and experienced users. Through open coding, we identified the following major themes related to this aspect.
    \begin{enumerate}
        \item \textbf{Customization} (40 issues, or 36.0\%): Many issues of this usability aspect centered around designing or building new features for users to alter the system's behavior or appearance in order to satisfy their personalized needs. For example, in Google Colab issue \#533, \textit{\enquote{Allow changing font size and picking font type inside the code cell}}, the participants required new features to customize font.
        \item \textbf{Extendability/interoperability/alternatives} (31 issues, or 27.92\%): One aspect of software development is its ability to be extended with additional features and/or alternatives to facilitate end users. Besides, this theme covers the subjects describing the design of software for working on different operating systems. For example, Jupyter Lab issue \#903, \textit{\enquote{Add prompt div for rendered markdown cells}}, requested modifying the current code structure in order to support extensions.
        \item \textbf{Shortcuts} (29 issues, or 26.1\%): Issues of this usability aspect are also frequently related to the features causing a quicker and more direct way of using the application. This theme is not only related to keyboard shortcuts but also concerns UI components (e.g., buttons) that serve as a shortcut. Apart from that, this theme focuses on making UI controls or related components easy to access for the users. For example, in CoCalc issue \#1272, the issue poster requested a button for easily exporting files.
        \item \textbf{Multitasking} (8 issues, or 7.2\%): This theme concentrates on features allowing users to do multiple things simultaneously. For example, CoCalc issue \#478, \textit{\enquote{popout current tab to window}}, proposes a new tab/window management approach to support multitasking.
        \item \textbf{Context awareness} (2 issues, or 1.8\%): This theme is related to allowing users to be aware of the context in order to speed up their tasks. For example, Jupyter Lab issue \#447, \textit{\enquote{Command palette interaction}}, reported a usability issue related to losing context.
        \item \textbf{Performance} (1 issue, or 0.9\%): This theme concentrates on factors contributing to the response speed and resource allocation of the applications. Particularly, Google Colab issue \#253 focuses on a request for an increase in RAM of Google Colab in order to improve performance.
    \end{enumerate}

\textbf{\#8: Aesthetic and minimalist design} (a total of 84 issues, or 27.5\% in the entire sample). The issues related to this usability aspect focus on suggestions and problems concerning visual design. We found the following major themes in these issues.
    \begin{enumerate}
        \item \textbf{Spacing/layout/format} (36 issues, or 42.8\%): This theme describes issues reporting the spacial arrangements of UI elements. An example of this theme is Cocalc issue \#4205, \textit{\enquote{Extra padding in latex warning/error tooltip boxes}}, which reports a visual formatting issue of a UI component.
        \item \textbf{Color/highlighting/theme} (14 issues, or 16.7\%): This theme focuses on color-related visual design issues of the project. For example, Jupyter Lab issue \#4867, \textit{\enquote{Dark theme scroll bar UI}}, reports problems related to the dark theme of the UI.
        \item \textbf{Fonts/texts} (12 issues, or 14.3\%): The issues of this theme concentrate on the display of textual content in the applications. One example of this is VSCode issue \#60965, \textit{\enquote{Font renders weird}}.
        \item \textbf{Icon/logo} (8 issues, or 9.5\%): This theme depicts the design issues related to the logos and icons in the projects. One example of this type of issue is VSCode issue \#11274, \textit{\enquote{The app logo is not decent}}.
        \item \textbf{Redundant or missing UI components} (7 issues, or 8.3\%): This theme is related to the problem of having an unneeded UI component or a needed UI component not showing. For example, Jupyter Lab issue \#5354, \textit{\enquote{Remove "Trust Notebook" menu item}}, suggested to remove an unnecessary UI component; and Atom issue \#15442, \textit{\enquote{Scroll bars not showing in center pane items}}, reported a missing component.
        \item \textbf{Size/resizing} (7 issues, or 8.3\%): This theme focuses on reporting the issues impacting the size of the elements or the need to resize them. For instance, VSCode issue \#8520, \textit{\enquote{Cutting off the error text}}, requested to resize the error text label.
    \end{enumerate}
    
\textbf{\#9: Help users recognize, diagnose, and recover from errors} (a total of 32 issues, or 10.5\% in the entire sample). This usability aspect concerns how well the error messages and suggested solutions can help users resolve the problem. The major themes we identified in the issues include:
    \begin{enumerate} 
        \item \textbf{Mysterious error} (11 issues, or 34.4\%): This type of issues expresses user interaction design problems that are difficult to understand or unexpectedly happen. For example, Atom issue \#11607, \textit{\enquote{Starting half window with specific project}}, reported an unexpected error without any user message.
        \item \textbf{Error message not helpful} (11 issues, or 34.4\%): In this type of issues, the participants suffered from a situation where the error message did not provide helpful and/or practical information.       One example of this is Jupyter Lab issue \#5294, \textit{\enquote{jupyter notebook not working}}, concentrating on not having a helpful message to guide users to fix the problem.
        \item \textbf{Other features for supporting error recognition and recovery} (10 issues, or 31.2\%): This theme includes issues discussing solutions different than error messages for helping users recognize and recover from errors. An example is Cocalc issue \#1032, \textit{\enquote{when a latex file has an error, put a button right in the preview pane, and a remark that you must fix this error to see the rest of the document}}.
    \end{enumerate}

\textbf{\#1: Visibility of system status} (a total of 28 issues, or 9.2\% in the entire sample). This usability aspect is related to interaction design concerns that allow users to see and be updated with what is going on with the system. We identified the following themes in the issues.
    \begin{enumerate}
        \item \textbf{Visibility of UI components related to status representation} (14 issues, or 50.0\%): This theme discusses the mechanism of displaying/hiding user interaction design elements that represent system status. Sometimes the disappearance of design elements leads to usability and interaction issues. Atom issue \#11681, \textit{\enquote{Bookmark icons should stay visible when line numbers gutter is hidden}} is one example of this type of issues. 
        \item \textbf{User message} (14 issues, or 50.0\%): This theme is related to issues that describe problems of missing suitable messages to update users about the situation and condition of the system. For instance, an end user of Google Colab posted Google Colab issue \#951, \textit{\enquote{Google Colab is not connecting to Google Drive \#URGENT \#URGENT \#URGENT}}, has not known the current system status in order to take steps.
    \end{enumerate}

\textbf{\#4: Consistency and standards} (a total of 21 issues, or 6.9\% in the entire sample). This usability aspect highlights the consistency of factors such as terms, situations, or actions within the system, across different products, or in relation to standards. We identified the following themes in the sampled issues.
    \begin{enumerate} 
        \item \textbf{Do not work as expected} (18 issues, or 85.7\%): Users reported issues when they found that the system did not work as expected or the display was not consistent with their expectations. These expectations are often established from conventions or previous experience using other software. For example, the user reporting Jupyter Lab issue \#2346, \textit{\enquote{Pressing the up arrow in console should put the cursor at the end of the previous command}}, was expecting something else when using the console.
        \item \textbf{Follow the standards} (3 issues, or 14.3\%): This code captures issues requesting application features to follow certain standards. For example, the user reporting the Cocalc issue \#3249, \textit{\enquote{double quotes in comments in assignment grade should be doubled in exported csv}}, complained that the CSV file generator was not consistent with the standard.
    \end{enumerate}

\textbf{\#5: Error prevention} (a total of 14 issues, or 4.6\% in the entire sample). This usability aspect concerns the design to prevent the occurrence of problems in the first place.
    \begin{enumerate}
        \item \textbf{Not preventing accidents} (7 issues, or 50.0\%): This theme covers issues that should have been eliminated from the system to remove error-prone conditions and situations. One example of this category is \textit{VSCode issue \#88742, \enquote{Searching in quick pick is not reliable}}.
        \item \textbf{Fails of ``smart'' features} (5 issues, or 36.7\%): This theme covers issues focusing on reporting problems in the supposedly smart components of the system. These issues usually originated from not considering human control in the design of the automated features. An example is VSCode issue \#29978, \textit{\enquote{Css/Scss autocomplete gives wrong result the first time when you press Tab}}.
        \item \textbf{Action awareness} (2 issues, or 14.3\%): This theme covers issues highlighting the need for users to be more aware of their actions to prevent errors. As one example of this theme, Atom issue \#1708, \textit{\enquote{creating a folder is weird}}, discusses an issue in which the user did not understand what is expected by the system and made an error.
    \end{enumerate}

Our results also revealed that three usability aspects barely appeared in the usability issues sample. They are \textit{\#2: Match between system and the real world} (0.7\%), \textit{\#3: User control and freedom} (0.7\%), and \textit{\#6: Recognition rather than recall} (0.7\%). Particularly, heuristic \#3 is not discussed in the regular code editor projects at all and is only seen in the CoCalc project. Additionally, in our sampled issues from the data science notebook projects, Google Colab did not have any issue labelled by four usability aspects (i.e., heuristics \#2, \#3, \#6, \#10) and Jupyter Lab's issues were not annotated with three aspects (i.e., heuristics \#2, \#3, \#6). For the code editor projects, there was not any issue in VSCode identified as heuristics \#3 and \#6, while in Atom \#2 and \#3 were not labelled.

\vspace{4pt}
\begin{tcolorbox}[colframe=black,colback=gray!10,boxrule=0.5pt,arc=.3em,boxsep=-1mm]
\textbf{RQ2 Main Findings}: 
An average of 17.4\% issues across the five analyzed projects are related to usability. The most frequently discussed usability aspects are related to (1) flexibility and efficiency and (2) aesthetics; issues targeting these aspects constitute about 63.9\% of all usability issues in our samples.
\end{tcolorbox}

%% file: s_Results_RQ3.tex
\subsection{RQ3: Distinctive characteristics of usability issues}

We first analyzed the use of visual contents, the number of reactions, the sentiments, and the tones of usability versus non-usability issue posts, as well as in the issues coded with different usability aspects; this analysis is done on each project sample (see Section~\ref{sec:rq3-methods}). We found that in all projects, users tended to use significantly more visual content when posting usability issues (45.5\% usability issues were posted with visual content in Google Colab, 43.8\% in Cocalc, 45.5\% in Jupyter Lab, 43.2\% in Atom, and 54.2\% in VSCode) than when posting non-usability issues (16.1\%, 13.5\%, 27.0\%, 14.5\%, and 23.5\%, respectively; $p<0.001$). There is also a significant difference in using visual content among the ten usability aspects in all projects ($p<0.001$). Particularly, in all projects, usability issues touching the heuristic \textit{\#7: Flexibility and efficiency of use} are less likely to be posted with visual content (33.3\%, 19.0\%, 36.4\%, 8.3\%, and 39.4\%, respectively in the five projects; $p<0.01$) when compared to other usability aspects; at the same time, in Cocalc and VSCode, usability issues touching the heuristic \textit{\#8: Aesthetic and minimalist design} are significantly more likely to be posted with visual content (77.3\% and 69.0\%, respectively; $p<0.01$).

Because of the important role of visual content in usability issues, we further investigated the type of visual content used in issue posts and the purpose of using the visual content. This analysis is done using a previous research codebook created by Agrawal et al.~\cite{Agrawal_2022visual}, which classified visual content in OSS issue discussions into seven types and six purposes. By analyzing our sampled dataset using this framework, we found that the most frequently used visual content types are \textit{Main product UI} (33.1\% or 44 issues) and \textit{UI component highlight} (22.5\% or 30 issues), followed by \textit{IDE or browser developer tools} (12.8\%), \textit{Peripheral product UI} (12.0\%), \textit{Code snippets} (11.3\%), \textit{Local command-line terminal/console} (4.5\%), and \textit{System window} (3.8\%). Moreover, the visual contents were most frequently used for the purposes of \textit{Illustrating a problem} (76.7\% or 102 issues) and \textit{Proposing enhancement or new feature} (14.3\% or 19 issues), followed by \textit{Illustrating expected behavior} (6.0\%), \textit{Showing or testing an implemented feature} (2.3\%), \textit{Providing instructions or workarounds} (0.8\%) and \textit{Showing a fixed problem or new feature implemented} (0.8\%).

Regarding the communities' emoji reaction to the issue posts, for three projects (Google Colab, Jupyter Lab, and VSCode), we found that usability issues received a significantly larger number of reactions from the community (a mean of 1.09, 1.96, and 1.89 reactions per issue, for the three projects respectively) than non-usability issues (0.21, 0.44, and 0.62, respectively). This is not found in the other two projects. When examining the usability aspects captured by heuristics, we found that among issues in Google Colab and VSCode, those related to heuristic \textit{\#7: Flexibility and efficiency of use} have received a significantly larger number of reactions (2.17 and 2.85 per issue, respectively) than non-usability issues.

Concerning sentiments and tones, we did not find a significant difference between usability and non-usability issues. Overall, slightly more than half of the issues were identified with a neutral sentiment (\textit{cross-project average}~$=57.1\%$ and $SD=9.3\%$, with 55.1\% and 57.7\% for usability and non-usability issues, respectively); for the other half, a cross-project average of 26.6\% (28.3\% and 26.3\% for usability and non-usability issues, respectively) were identified with a positive sentiment and 16.3\% with a negative sentiment. The most frequent tone was \textit{sad} (\textit{cross-project average}~$=45.1\%$), followed by \textit{frustrated} (10.6\%) and \textit{polite} (10.0\%); at the same time, the appearance of satisfied, sympathetic, and impolite tones were very rare ($<0.3\%$).

Among the usability aspects captured by heuristics, the only significant results were found on the tones of issue posts in Cocalc and VSCode projects ($p<0.001$ and $p<0.05$, respectively); no differences were found among the sentiments. In Cocalc, issues related to \textit{\#7: Flexibility and efficiency of use} and \textit{\#8 Aesthetic and minimalist design} are significantly more likely (respectively 66.7\% and 50.0\% of the time, $p<0.05$) to be identified with a \textit{sad} tone than issues touching other usability aspects. In VSCode, issues related to \textit{\#8 Aesthetic and minimalist design} are more likely (41.3\%, $p<0.05$) to display a \textit{frustrated} tone, while issues related to \textit{\#7: Flexibility and efficiency of use} are more likely to display a \textit{sad} tone (36.4\%, $p<0.05$).

\vspace{4pt}
\begin{tcolorbox}[colframe=black,colback=gray!10,boxrule=0.5pt,arc=.3em,boxsep=-1mm]
\textbf{RQ3 Main Findings (a)}: 
Compared to non-usability issues, usability issue reports are more likely to include visual content and receive more emoji reactions from the community. Similar to non-usability issues, usability issues tend to display a neutral and positive sentiment with a sad, frustrated, and polite tone. 
\end{tcolorbox}

Then, we considered the discussion length, the number of participants, and three discussion timeline metrics to understand the characteristics of discussion threads of usability issues versus non-usability issues in the five projects, as well as among the issues touching different usability aspects (see Section~\ref{sec:rq3-methods}). In all five projects, ANOVA tests did not find a significant difference in the discussion length, measured by the number of comments, on both independent variables. However, in Jupyter Lab and VSCode, usability issues involved a significantly larger number of participants ($mean=3.1$ and 2.7, respectively) than non-usability issues ($mean=2.5$ and 2.1, $p<0.05$); but no significant differences were observed among issues related to different usability aspects.

Regarding the discussion timeline, we did not find significant differences in the community response time (i.e., time to the first comment) between usability and non-usability issues or among usability aspects in all five projects. However, we found that in Jupyter Lab and VSCode, usability issues took longer to receive the last comment ($mean=183.1$~days and $107.7$~days, respectively) and get closed ($mean=205.4$~days and $102.5$~days, respectively) than non-usability issues (time to last comment $mean=89.5$~days and $43.5$~days and time to issue close $mean=99.6$~days and $41.6$~days, respectively, $p<0.01$); but no significant difference was found among the usability aspects captured by heuristics. We also did not observe such differences in the other projects.

\vspace{4pt}
\begin{tcolorbox}[colframe=black,colback=gray!10,boxrule=0.5pt,arc=.3em,boxsep=-1mm]
\textbf{RQ3 Main Findings (b)}: Usability issues tended to receive a similar number of comments to non-usability issues. But in certain projects (Jupyter Lab and VSCode), usability issues involved more participants, who took longer to discuss and close the issues.
\end{tcolorbox}

%% file: s_Results_RQ4.tex
\subsection{RQ4: Experiences and backgrounds of usability issue posters}

\begin{table*}[t]
\centering
\caption{Summary of statistical test results on the experiences of the three groups of issue posters. (*** p<0.001, ** p<0.01, * p<0.05)}
\small
\resizebox{\textwidth}{!}{
\begin{tabular}{ccc|cc|cc|cc|cc|cc}
\hline
 &  &  & \multicolumn{2}{c|}{Google Colab} & \multicolumn{2}{c|}{Cocalc} & \multicolumn{2}{c|}{Jupyter Lab} & \multicolumn{2}{c|}{Atom} & \multicolumn{2}{c}{VSCode} \\
Metrics & Group (I) & Group (J) & Mean (I) & Diff (I-J) & Mean (I) & Diff (I-J) & Mean (I) & Diff (I-J) & Mean (I) & Diff (I-J) & Mean (I) & Diff (I-J) \\\hline

\multirow{6}{*}{\# Posted Issues} & \multirow{2}{*}{Uposters} & NUPosters & \multirow{2}{*}{1.24} & -5.65 & \multirow{2}{*}{46.75} & 34.44 & \multirow{2}{*}{5.78} & 1.48 & \multirow{2}{*}{7.21} & 1.21 & \multirow{2}{*}{26.25} & -31.02 \\
 &  & BothPosters &  & -8.89 &  & -140.77 &  & -109.04 ** &  & -239.50 ** &  & -1140.85 *** \\
 & \multirow{2}{*}{NUPosters} & Uposters & \multirow{2}{*}{6.89} & 5.65 & \multirow{2}{*}{12.31} & -34.44 & \multirow{2}{*}{4.30} & -1.48 & \multirow{2}{*}{6.00} & -1.21 & \multirow{2}{*}{57.27} & 31.02 \\
 &  & BothPosters &  & -3.24 &  & -175.21 *** &  & -110.52 *** &  & -240.71 *** &  & -1109.83 *** \\
 & \multirow{2}{*}{BothPosters} & Uposters & \multirow{2}{*}{10.13} & 8.89 & \multirow{2}{*}{187.52} & 140.77 & \multirow{2}{*}{114.82} & 109.04 ** & \multirow{2}{*}{246.71} & 239.50 ** & \multirow{2}{*}{1167.10} & 1140.85 *** \\
 &  & NUPosters &  & 3.24 &  & 175.21 *** &  & 110.52 *** &  & 240.71 *** &  & 1109.83 *** \\[3pt]\hline

\multirow{6}{*}{\# Posted Comments} & \multirow{2}{*}{Uposters} & NUPosters & \multirow{2}{*}{1.24} & -13.13 & \multirow{2}{*}{231.25} & 81.11 & \multirow{2}{*}{32.52} & 10.02 & \multirow{2}{*}{93.17} & 56.22 & \multirow{2}{*}{73.09} & -289.95 \\
 &  & BothPosters &  & -51.32 &  & -196.12 &  & -527.70 *** &  & -1228.40 ** &  & -10614.81 ** \\
 & \multirow{2}{*}{NUPosters} & Uposters & \multirow{2}{*}{14.37} & 13.13 & \multirow{2}{*}{150.14} & -81.11 & \multirow{2}{*}{22.50} & -10.02 & \multirow{2}{*}{36.95} & -56.22 & \multirow{2}{*}{363.04} & 289.95 \\
 &  & BothPosters &  & -38.19 &  & -277.23 &  & -537.72 &  & -1284.62 *** &  & -10324.86 *** \\
 & \multirow{2}{*}{BothPosters} & Uposters & \multirow{2}{*}{52.56} & 51.32 & \multirow{2}{*}{427.37} & 196.12 & \multirow{2}{*}{560.22} & 527.70 *** & \multirow{2}{*}{1321.57} & 1228.40 ** & \multirow{2}{*}{10687.90} & 10614.81 ** \\
 &  & NUPosters &  & 38.19 &  & 277.23 &  & 537.72 &  & 1284.62 *** &  & 10324.86 *** \\[3pt]\hline

\multirow{6}{*}{\# Years Joined} & \multirow{2}{*}{Uposters} & NUPosters & \multirow{2}{*}{5.84} & 0.99 & \multirow{2}{*}{8.75} & 1.34 & \multirow{2}{*}{7.30} & 0.86 & \multirow{2}{*}{8.93} & 0.90 & \multirow{2}{*}{7.60} & 1.23 * \\
 &  & BothPosters &  & -0.41 &  & -0.57 &  & -2.09 *** &  & -2.21 * &  & -0.50 \\
 & \multirow{2}{*}{NUPosters} & Uposters & \multirow{2}{*}{4.85} & -0.99 & \multirow{2}{*}{7.41} & -1.34 & \multirow{2}{*}{6.44} & -0.86 & \multirow{2}{*}{8.03} & -0.90 & \multirow{2}{*}{6.37} & -1.23 * \\
 &  & BothPosters &  & -1.40 &  & -1.91 &  & -2.95 *** &  & -3.11 *** &  & -1.73 * \\
 & \multirow{2}{*}{BothPosters} & Uposters & \multirow{2}{*}{6.25} & 0.41 & \multirow{2}{*}{9.32} & 0.57 & \multirow{2}{*}{9.39} & 2.09 *** & \multirow{2}{*}{11.14} & 2.21 & \multirow{2}{*}{8.10} & 0.50 \\
 &  & NUPosters &  & 1.40 &  & 1.91 &  & 2.95 *** &  & 3.11 *** &  & 1.73 * \\[3pt]\hline

\multirow{6}{*}{\# Repositories} & \multirow{2}{*}{Uposters} & NUPosters & \multirow{2}{*}{36.12} & 2.17 & \multirow{2}{*}{130.50} & 93.78 & \multirow{2}{*}{87.70} & 41.86 * & \multirow{2}{*}{65.83} & 11.24 & \multirow{2}{*}{45.72} & 5.85 \\
 &  & BothPosters &  & -21.69 &  & 66.66 &  & -61.26 * &  & -26.74 &  & -53.58 ** \\
 & \multirow{2}{*}{NUPosters} & Uposters & \multirow{2}{*}{33.95} & -2.17 & \multirow{2}{*}{36.72} & -93.78 & \multirow{2}{*}{45.84} & -41.86 * & \multirow{2}{*}{54.59} & -11.24 & \multirow{2}{*}{39.87} & -5.85 \\
 &  & BothPosters &  & -23.86 &  & -27.12 &  & -103.12 *** &  & -37.98 &  & -59.43 *** \\
 & \multirow{2}{*}{BothPosters} & Uposters & \multirow{2}{*}{57.81} & 21.69 & \multirow{2}{*}{63.84} & -66.66 & \multirow{2}{*}{148.96} & 61.26 * & \multirow{2}{*}{92.57} & 26.74 & \multirow{2}{*}{99.30} & 53.58 ** \\
 &  & NUPosters &  & 23.86 &  & 27.12 &  & 103.12 *** &  & 37.98 &  & 59.43 ***\\[3pt]\hline
\end{tabular}
}
\label{tab:rq2-results}
\end{table*}

\begin{table*}[t]
\centering
\caption{Summary of statistical test results on the roles of the three groups of issue posters. (*** p<0.001, ** p<0.01, * p<0.05)}
\small
\resizebox{\textwidth}{!}{
\begin{tabular}{ccc|cc|cc|cc|cc|cc}
\hline
 &  &  & \multicolumn{2}{c|}{Google Colab} & \multicolumn{2}{c|}{Cocalc} & \multicolumn{2}{c|}{Jupyter Lab} & \multicolumn{2}{c|}{Atom} & \multicolumn{2}{c}{VSCode} \\
User Role & Group (I) & Group (J) & \# (I) & Diff (I-J) & \# (I) & Diff (I-J) & \# (I) & Diff (I-J) & \# (I) & Diff (I-J) & \# (I) & Diff (I-J) \\\hline

\multirow{6}{*}{Member/Collaborator} & \multirow{2}{*}{Uposters} & NUPosters & \multirow{2}{*}{0} & -1 & \multirow{2}{*}{0} & 0 & \multirow{2}{*}{2} & -15 *** & \multirow{2}{*}{1} & -5 & \multirow{2}{*}{0} & -18 *** \\
 &  & BothPosters &  & 0 &  & -3 &  & -8 *** &  & -3 &  & -12 *** \\
 & \multirow{2}{*}{NUPosters} & Uposters & \multirow{2}{*}{1} & 1 & \multirow{2}{*}{0} & 0 & \multirow{2}{*}{17} & 15 *** & \multirow{2}{*}{6} & 5 & \multirow{2}{*}{18} & 18 *** \\
 &  & BothPosters &  & 1 &  & -3 &  & 7 &  & 2 &  & 6 \\
 & \multirow{2}{*}{BothPosters} & Uposters & \multirow{2}{*}{0} & 0 & \multirow{2}{*}{3} & 3 & \multirow{2}{*}{10} & 8 *** & \multirow{2}{*}{4} & 3 & \multirow{2}{*}{12} & 12 *** \\
 &  & NUPosters &  & -1 &  & 3 &  & -7 *** &  & -2 &  & -6 \\[3pt]\hline
 
\multirow{6}{*}{Contributor} & \multirow{2}{*}{Uposters} & NUPosters & \multirow{2}{*}{0} & -1 & \multirow{2}{*}{0} & -3 & \multirow{2}{*}{13} & -2 & \multirow{2}{*}{2} & -25 *** & \multirow{2}{*}{4} & -13 * \\
 &  & BothPosters &  & -3 &  & -14 *** &  & 1 &  & -4 &  & -4 \\
 & \multirow{2}{*}{NUPosters} & Uposters & \multirow{2}{*}{1} & 1 & \multirow{2}{*}{3} & 3 & \multirow{2}{*}{15} & 2 & \multirow{2}{*}{27} & 25 *** & \multirow{2}{*}{17} & 13 * \\
 &  & BothPosters &  & -2 &  & -11 * &  & 3 &  & 21 *** &  & 9 \\
 & \multirow{2}{*}{BothPosters} & Uposters & \multirow{2}{*}{3} & 3 & \multirow{2}{*}{14} & 14 *** & \multirow{2}{*}{12} & -1 & \multirow{2}{*}{6} & 4 & \multirow{2}{*}{8} & 4 \\
 &  & NUPosters &  & 2 &  & 11 * &  & -3 &  & -21 *** &  & -9 \\[3pt]\hline

\multirow{6}{*}{Other} & \multirow{2}{*}{Uposters} & NUPosters & \multirow{2}{*}{25} & -207 *** & \multirow{2}{*}{4} & -22 *** & \multirow{2}{*}{25} & -102 *** & \multirow{2}{*}{26} & -241 *** & \multirow{2}{*}{64} & -143 *** \\
 &  & BothPosters &  & 12 &  & 0 &  & 7 &  & 22 *** &  & 64 *** \\
 & \multirow{2}{*}{NUPosters} & Uposters & \multirow{2}{*}{232} & 207 *** & \multirow{2}{*}{26} & 22 *** & \multirow{2}{*}{127} & 102 *** & \multirow{2}{*}{267} & 241 *** & \multirow{2}{*}{207} & 143 *** \\
 &  & BothPosters &  & 219 *** &  & 22 *** &  & 109 *** &  & 263 *** &  & 207 *** \\
 & \multirow{2}{*}{BothPosters} & Uposters & \multirow{2}{*}{13} & -12 & \multirow{2}{*}{4} & 0 & \multirow{2}{*}{18} & -7 & \multirow{2}{*}{4} & -22 *** & \multirow{2}{*}{0} & -64 *** \\
 &  & NUPosters &  & -219 *** &  & -22 *** &  & -109 *** &  & -263 *** &  & -207 *** \\[3pt]\hline
\end{tabular}%
}
\label{tab:rq2-results-roles}
\end{table*}

To answer RQ4, we considered three groups of issue posters: (1) those who only posted usability issues (\textit{UPosters}), (2) those who only posted non-usability issues (\textit{NUPosters}), and (3) those who posted both types of issues (\textit{BothPosters}). We compared these three groups in each project on the five metrics (dependent variables) summarized in Section~\ref{sec:rq2-methods}. The statistical test results are summarized in Table~\ref{tab:rq2-results} and Table~\ref{tab:rq2-results-roles}. We found that in many projects (particularly Jupyter Lab, Atom, and VSCode), issue posters who have posted both usability issues and non-usability issues, when compared to those who have only posted usability issues or only posted non-usability issues, have (1) created a significantly larger amount of issues, (2) made significantly more comments in issue discussions, (3) joined GitHub for a significantly longer period of time, and (4) owned a larger number of repositories. Interestingly, we also found that in VSCode, users who only posted usability issues have joined GitHub for a longer period of time than those who only posted non-usability issues; and in Jupyter Lab, usability issue posters have owned more repositories those who only posted non-usability issues.

When examining the role composition of these issue posters, we found that in Jupyter Lab and VSCode, there was a significantly larger number of \textit{Members/Collaborators} who posted both usability and non-usability issues or only non-usability issues than those who only posted usability issues. Moreover, among the \textit{Other} participants, a significantly larger number only posted non-usability issues, followed by usability-only posters and those who posted both types of issues; this is observed in all five projects. For code \textit{Contributors}, however, there is a different trend across the five projects. In Cocalc, there was a significantly larger number of \textit{Contributors} who posted both usability and non-usability issues, when compared to the two other groups. On the other hand, in Atom and VSCode, code \textit{Contributors} tended to be those who only posted non-usability issues.

\vspace{4pt}
\begin{tcolorbox}[colframe=black,colback=gray!10,boxrule=0.5pt,arc=.3em,boxsep=-1mm]
\textbf{RQ4 Main Findings}: 
Users who have posted both usability and non-usability issues were generally more active in issue discussions and had more experiences. Usability issue posters were equally active compared to non-usability issue posters and can have more experience in certain projects (Jupyter Lab and VSCode). Usability issues were posted by all user roles.
\end{tcolorbox}

%% file: s_Discussion.tex
Through this study, we provided a rather comprehensive profile characterizing usability issues in data science notebook and code editor OSS projects. In this section, we discuss the major takeaways of our findings, as well as their implications for OSS practitioners and researchers who focused on promoting collaboration on niche topics such as OSS usability.

\subsection{Synthesizing the key results}

\subsubsection{Usability issues are extensively discussed in the ITS, actively involving the community}
Across the OSS projects we investigated, between 9.9\% and 25.1\% of the issues are related to usability, with an average ratio of 17.4\%. This seems to be a small number. But considering the large range of topics discussed in the ITS, the focus on usability in those projects is, in fact, substantial. This echoes our topic analysis results, in which seven out of the 26 distilled topics were related to usability. Moreover, through our quantitative analysis, we found that usability issues attracted more emoji reactions and tended to involve more (and potentially more diverse) participants. These results indicate that ITS users are motivated to contribute to the usability issue discussions. We also identified that experienced GitHub users contributed more to usability issues -- a long engagement with OSS communities may have allowed them to be more interested in and aware of usability-related topics. Together, our results provided direct evidence that usability issues are extensively targeted and discussed in the ITS and are of avid interest to the OSS communities. This finding supports the previous studies that argued for the importance of ITSs for OSS usability~\cite{wang2020argulens, bach2009floss, Wang2022IEEESoftware, terry2010perceptions}.

\subsubsection{Current OSS usability issues are mostly focused on efficiency and aesthetics}
Although the ITS is extensively used for raising and addressing usability issues, the scope of usability discussion seems to be limited. Through topic modeling, we found that the usability-related topics were mostly focused on visual design and user interaction mechanisms. When manually analyzing the usability aspects touched by the issues, we also found that there were several dominant aspects of usability appeared in the issue discussions. Particularly, aspects related to (\textit{Flexibility and efficiency of use}) and (\textit{Aesthetic and minimalist design}) constituted more than 60\% of the usability issues. On the other hand, several important usability aspects were rarely targeted, including (\textit{User control and freedom}), (\textit{Match between system and the real world}), and (\textit{Recognition rather than recall}). We speculate that this is partially due to the system-centric and post-hoc perspective usually taken by OSS communities towards usability~\cite{Wang2022IEEESoftware}: efficiency issues (e.g., customization, extendability, and shortcuts, as we identified through open coding) and aesthetic issues (e.g., layout, color, and fonts) tend to be system characteristics that can be readily discovered and addressed by developers once a version of the system is implemented, while other usability aspects may require deeper, upfront analysis and UI/UX design. These results indicated a need to support awareness of diverse usability concerns and broaden the spectrum of usability discussions in the OSS communities. 

\subsubsection{Collaborative visual communication is important but complex for OSS usability issues}
Our findings revealed that usability issues were more frequently reported with visual support, indicating a desire for visual communications in this collaborative endeavor. While previous work revealed a need for visual discussion in programming forums~\cite{Zhu2015}, our study highlighted the prominence of this need in usability discussions. Specifically, we found that visual content was frequently used to represent the main UI or UI components, in order to \textit{illustrate a problem} or \textit{propose enhancement/new features}. Interestingly, issues related to aesthetics were more frequently posted with the support of visual content, while issues related to efficiency tended to be posted with only textual content. At the same time, we identified that usability issues involved OSS community members who assumed various roles (including core members/collaborators, code contributors, and those who only contributed to issue discussions). Together, these factors indicate that the need for collaborative visual communication can differ depending on the aspect that the issue touches on and the users who engage in the discussion. However, current ITSs typically do not provide sufficient support for visual communication and collaboration, if at all. On the other hand, design tools with collaboration features such as Figma\footnote{https://www.figma.com} and InVision\footnote{https://www.invisionapp.com} are usually too heavy-weight for OSS usability issue discussion and may not be easily adopted by OSS communities. Investigating methods and tools that support lightweight visual collaboration in the context of OSS usability may help achieve a desirable balance.

\subsection{Implications to practice and research}
Based on the results and takeaways generated from our study, we propose some concrete implications to practice and research that aim to promote OSS usability. These implications are summarized in Table~\ref{tab:implications} and described below.

\begin{table*}[t]
\caption{Implications of our results and takeaways to OSS practitioners and researchers.}
\label{tab:implications}
\centering
\small
\resizebox{\textwidth}{!}{
\begin{tabular}{p{4cm}p{5.5cm}p{5.5cm}}
    \toprule
    \textbf{Key results and takeaways} & \textbf{Implications for practitioners} & \textbf{Implications for researchers}\\
    
    \midrule
    Usability issues are extensively discussed in the ITS, actively involving the community & 
    \begin{minipage}[t]{\linewidth}
    \begin{itemize}[leftmargin=*]
        \item Facilitate cross-stakeholder communication and collaboration
        \item Use usability heuristics as a framework to distinguish usability issues from non-usability issues
    \end{itemize}
    \end{minipage} &
    \begin{minipage}[t]{\linewidth}
    \begin{itemize}[leftmargin=*]
        \item Investigate automated techniques to detect usability issues 
        \item Explore ways to support inexperienced users in usability issue discussions
    \end{itemize}
    \end{minipage}\\

    \\
    Current OSS usability issues are mostly focused on efficiency and aesthetics &
    \begin{minipage}[t]{\linewidth}
    \begin{itemize}[leftmargin=*]
        \item Raise awareness of usability issues beyond efficiency and aesthetics, leveraging usability heuristics
    \end{itemize}
    \end{minipage} &
    \begin{minipage}[t]{\linewidth}
    \begin{itemize}[leftmargin=*]
        \item Address barriers that hinder OSS communities to consider a wider range of usability aspects
    \end{itemize}
    \end{minipage}\\

    \\
    Collaborative visual communication is important but complex &
    \begin{minipage}[t]{\linewidth}
    \begin{itemize}[leftmargin=*]
        \item Encourage discussions of usability issues to be framed around visual content
    \end{itemize}
    \end{minipage} &
    \begin{minipage}[t]{\linewidth}
    \begin{itemize}[leftmargin=*]
        \item Explore techniques and tools that support lightweight visual collaboration
    \end{itemize}
    \end{minipage}\\
    \bottomrule
\end{tabular}
}
\end{table*}

\subsubsection{Implications for practitioners}
We identified the following implications of our results for OSS practitioners to better address usability issues.

\textbf{Facilitate cross-stakeholder communication and collaboration in ITSs.}
Our results revealed that ITSs are still important for members of OSS communities, who have diverse backgrounds and roles, to collaboratively discuss and address usability issues. Thus investigating ways to facilitate cross-stakeholder communication, collaboration, and community building in ITSs to support OSS usability issue discussions will have practical impacts. Concretely, contributions to usability discussion need to be acknowledged and valued. The underrepresented stakeholders, such as end-users and designers, need to be encouraged to participate in this collaborative effort. Additionally, the discussion environment needs to be made friendly and free from toxicity and incivility~\cite{ferreira2021shut}.

\textbf{Use usability heuristics to distinguish usability issues from non-usability issues.}
A major problem of the current ITSs is that usability issues are mixed up together with many other topics, making them difficult to be discerned. Previous work has established the importance of distinguishing usability issues for them to be properly addressed~\cite{Wang2022IEEESoftware}. Our analysis demonstrated that identifying usability aspects using the well-known Nielsen's heuristics is a viable way to separate usability issues with manageable manual efforts. Thus, before a reliable automated technique is developed, OSS practitioners can leverage usability heuristics to classify usability issues.

\textbf{Raise awareness of usability issues beyond efficiency and aesthetics.}
We found that OSS communities rarely discuss usability aspects other than efficiency and aesthetics. This may have been affected by the current OSS culture. However, we argue that a consideration of diverse usability aspects that touch deeper user interaction issues can be a first step for an OSS community to pay more attention to this important problem. Thus, finding ways to educate OSS communities on the broad spectrum of usability aspects can have great value. We have already started to see such an effort in OSS organizations (such as Open Source Design\footnote{https://opensourcedesign.net}) and practitioner conferences (such as the design and usability tracks in FOSDEM\footnote{https://fosdem.org}).

\textbf{Encourage discussions of usability issues to be framed around visual content.}
We found that visual content plays an important role in the collaborative effort of discussing and addressing OSS usability issues. Previous work has indicated that using visual content in issues allows more OSS community members to participate in the discussion and is associated with richer discussion threads~\cite{Agrawal_2022visual}. Thus to facilitate diverse participants to engage in usability issue discussions, OSS communities can make efforts to encourage visual content use (e.g., in their contribution guidelines or issue templates). OSS tool developers also should consider this factor to provide features to facilitate visual communication. The GitLab Design Management feature\footnote{https://docs.gitlab.com/ee/user/project/issues/design\_management.html} can be considered as an early attempt to address this. However, the dynamics of the UI behaviors are often more difficult to communicate~\cite{Ozenc2010}; the design of OSS tools for usability could be inspired by related work about visual communication of interactive UI behaviors (e.g.,~\cite{Chen2021}).

\subsubsection{Implications for researchers} Our results also have the following implications for researchers who are focused on OSS and usability.

\textbf{Investigate automated techniques to detect usability issues.}
Automatically detecting usability issues from the large amount of issue threads can further motivate and facilitate addressing OSS usability. As an exploration, we investigated how well the topic models obtained from our study can be used to identify usability versus non-usability issues. For this, we rebuilt the topic models on the dataset without the samples that we manually labeled and then applied the models to the sampled data. We found that the topic models can help identify non-usability issues with satisfactory performance (F-1 score for notebook projects is 0.80 and F-1 score for code editor projects is 0.78), but with an inferior performance identifying usability issues (F-1 scores are 0.22 and 0.20, respectively). This indicates that an unsupervised technique such as topic modeling can help narrow down usability issues (by filtering out non-usability issues) for manual inspection. Additionally, our manually labeled dataset available in our replication package can be used to train supervised methods for automated usability issue detection.

\textbf{Explore ways to support inexperienced users in usability issue discussions}
Our results indicated that, on the one hand, tools supporting OSS usability collaboration should consider the needs of experienced users and help them manage and keep track of the many usability-related items they contribute to. On the other hand, end-users and designers are usually inexperienced in OSS platforms and desire alternative ways to be involved in usability issue discussion and collaboration~\cite{Hellman2021Facilitating}. To address this inclusiveness issue, future research is needed to investigate novel tools and techniques that integrate (1) the system-centric perspectives of the current OSS culture and (2) the needs of the underrepresented stakeholders such as end-users and designers.

\textbf{Address barriers that hinder OSS communities to consider a wider range of usability aspects.}
While we have identified that OSS communities tend to have a limited scope on usability aspects, future research is needed to address the factors that contributed to this phenomenon. Particularly, we currently do not fully understand the barriers that have prevented OSS communities from considering a wider range of usability aspects. Identifying and finding ways to break through those barriers will greatly promote OSS usability.

\textbf{Explore techniques and tools that support lightweight visual collaboration.}
Our results highlighted a need for lightweight visual collaboration in the discussion of OSS usability issues. While various designer-oriented tools exist (e.g., Figma and InVision) for collaboration oriented around visual artifacts, those tools do not fully address the particular characteristics of OSS communities. Future research should be first conducted to clarify the need to use visual content for different purposes, by different user roles, in the OSS context. Then, specific techniques and tools can be provided to satisfy those needs.

%% file: s_Limitations.tex
\section{Threads to validity and limitations}
Our study has the following major limitations. First, we were only able to analyze five OSS projects in our study in two application domains, data science notebooks and code editors. However, these projects were carefully selected to represent popular and active OSS applications and generated a large amount of issue discussion data for analysis. Albeit, future work needs to be conducted to establish the generalizability of our results. 

Second, in this project, we focused on established OSS communities around popular projects. This decision was made mainly because the usability of these projects often has a stronger impact on both, immediately, a large number of users and, indirectly, the broader field connected to the projects such as data science and software development. However, the usability of smaller OSS projects should not be ignored and needs to be addressed in future work.

Third, the topic modeling technique has innate limitations. Particularly, the topics were only identified based on word patterns and did not incorporate the real meanings of the words. Thus common words inevitably led to topics overlapping, even if the words had different meanings in different contexts. We thus only used topic modeling as a preliminary exploration of the general concerns discussed in the ITS and did not intend to use it for identifying usability issues. Additionally, the topic labels were created based on our interpretations. To minimize this threat, we used the entire cluster of keywords as well as the dominant issues in each topic to support our interpretation. The two authors also discussed the labeling results extensively. This methodology has been applied in other related works (e.g.,~\cite{bagherzadeh2019going}).

Fourth, our qualitative analysis and labeling are done on a sample of the entire dataset and may be subject to personal bias. To address this threat, we first ensured that we created a statistically representative sample of the dataset for manual analysis. Additionally, we followed a rigorous content analysis process that involved two coders in evaluating reliability and addressing disagreements.

Fifth, our quantitative analysis relies on natural language processing tools such as language detection, sentiment analysis, and tone analysis. These automated tools may not report accurate results. To mitigate this risk, for language detection, we selected the best-performing tool by comparing the tool prediction with manually labeled data. For sentiment and tone detection tools, we relied on a previous extensive investigation~\cite{Sanei2021impacts}.

Finally, this study is based on an analysis of artifacts created by OSS community members. Although this method provided rich information about the main focuses of the OSS communities regarding usability, it can only give vague hints about personal cognitive aspects such as goals, motivations, and challenges of usability issue posters. While some existing work has touched on these aspects~\cite{Wang2022IEEESoftware, terry2010perceptions}, more user studies with OSS community members need to be done in the future to help triangulate our findings.

%% file: s_Conclusion.tex
In this paper, we accumulated knowledge about the characteristics of usability issues that are discussed in the issue tracking system of open source software projects. Toward this end, we conducted an in-depth analysis of issues from five popular and active data science notebook and code editor projects. Our results revealed that while usability issues are extensively discussed in the issue tracking system, their scope tends to be limited to efficiency and aesthetics. We also found that usability issues had certain distinctive characteristics, incorporating more visual communication and involving community members with various experiences and roles. Overall, our study demonstrates a systematic characterization of OSS usability issues. Our results and their implications can inform OSS practices to facilitate collaboration among diverse stakeholders in usability issue discussions and inspire future research to address collaboration issues on niche topics in diverse communities. Eventually, we hope our efforts can contribute to the collective endeavour to create better, usable OSS applications.